\documentclass[aps,pre,twocolumn, showkeys]{revtex4-1}
\usepackage{ucs}
\usepackage[utf8x]{inputenc}
\usepackage{amsfonts}
\usepackage{amsmath}
\usepackage[american]{babel}
\usepackage[T1]{fontenc}
\usepackage{epsfig}
\usepackage{graphicx}
\usepackage{bm}
\usepackage{mathtools,amssymb,amsthm}
\usepackage{textcomp}
\usepackage{epsfig}
\usepackage{comment}

\date{\today}

\begin{document}
\author{T. M\"akinen}
\author{A. Miksic}
\author{M. Ovaska}
\author{Mikko J. Alava}
\affiliation{COMP center of excellence, Department of Applied Physics, Aalto University School of Science, P.O.Box 11100, FI-00076 AALTO, Espoo, Finland}

\title{Avalanches in wood compression}

\begin{abstract}
Wood is a multi-scale material exhibiting a complex viscoplastic response. We study avalanches in small wood samples in compression. "Woodquakes" measured by acoustic emission are surprisingly 
similar to earthquakes and crackling noise in rocks and laboratory tests on brittle materials. Both the distributions of event energies and of waiting (silent) times follow power-laws. The stress-
strain response exhibits clear signatures of localization of deformation to "weak spots" or softwood layers, as identified using Digital Image Correlation. Even though material structure-dependent 
localization takes place, the avalanche behavior remains scale-free.
\end{abstract} 
\maketitle

Materials do not deform smoothly in general which may be seen on many scales. Avalanches in fracture, interface dynamics of domain walls and irreversible plastic deformation exhibit power-law 
statistical properties and are thus "scale-free" \cite{Bonamy-2011, Alava-2006}. The intermittency of deformation or such non-linear response is a signature for collective phenomena 
\cite{Sethna-2001} and is important in practice to eventually predict the failure. Fracture avalanches are established in many materials \cite{Bonamy-2009, Sethna-2001}, mostly by acoustic emission 
(AE) technique \cite{Scholz-1968}-\cite{SilicaCeramics-2014}, and analogies are often drawn to "laboratory-scale earthquakes". Apart from the presence of power laws in the AE statistics, what other 
universal features can be distinguished is not clear. Theory suggests \cite{Alava-2006} why avalanches with the observed statistical properties exist, but hints about the relevance of the detailed 
physics, including the loading mode, the possible difference between 2D and 3D materials, the existence of a dominating crack and finally the material properties, like disorder and rheology.

Compressional loading is particularly interesting. When the material is crushed, the damaged structure and "debris" maintain the ability to carry the load via the frictional contacts, and 
deformation eventually localizes into shear bands. Such laboratory experiments may also serve as scaled-down models of earthquake faults. Recent research concentrates on signatures of criticality 
from models and the role of self-organization of deformation and load \cite{Girard-2012,Kun-2014}. Experimental studies focus on testing the analogy of brittle fracture of rocks 
\cite{Scholz-1968}-\cite{RockAE-DIC-2011} and heterogeneous \cite{Precursors-1998}-\cite{PapAE-14} or porous media \cite{Deschanel-2009}-\cite{SilicaCeramics-2014} with earthquakes 
\cite{Earthq-1999, Earthq-2009}. Data from small scales presents statistical laws like those from earthquake activity. An important question across the scales is whether (possible) correlations 
in crackling noise and deformation help to forecast large events or the final failure \cite{Precursors-1998, Precursors-2002, PapAE-08}. 

Many porous materials have a complex structure and response to compression. In this Letter, we deal with wood, a ubiquitous biological material with a cellular structure with hierarchical ordering. 
The wood cell arrangement depends on the annual growth cycle, leading to an alternating series of softwood and hardwood layers \cite{WoodMat-1999, WoodMat-2011}. The mechanical properties vary 
according to the species and origins of the original log and size of the sample, and the testing way (along the grains, etc.) \cite{WoodMat-2011, Wood-2013}. Here, we show the presence of avalanches 
in wood compression deformation. Acoustic and optical measurements together with strain/stress -curves demonstrate the existence of avalanches with scale-free properties, in this "non-brittle" 
material with complex structure. Earlier studies of AE in wood fracture exist for many applied purposes \cite{WoodAE-00}-\cite{WoodAE-DIC-13}, but not for the general physics.

\begin{figure}
\includegraphics[width=\columnwidth]{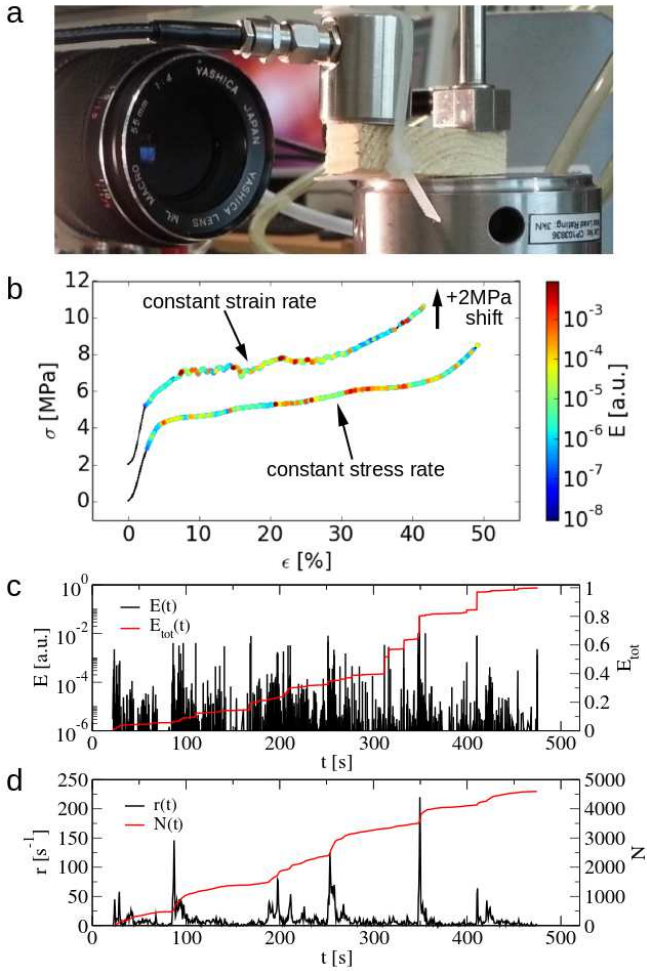}
\caption{\label{fig:1} a) Experimental set-up: One end of a wood piece is compressed in the machine by the piston extension stick. A piezoelectric sensor attached to the other end detects acoustic 
emission. A camera takes pictures of one sample side, enlightened with a LED lamp. b) Compression strain-stress curves for two typical experiments at constant strain- and stress-rate with AE energies 
represented with color-code; the curve for constant strain-rate test is shifted by +2MPa to avoid curve overlap. c) AE events (left y-axis) and cumulative AE energy normalized by the total energy 
(right y-axis) against time for constant strain rate compression. d) AE event rate and total number (left and right y-axis respectively) against time, during the same experiment.}
\end{figure}

\textit{Experiments -} Small pieces of dry pine (Pinus sylvestris) wood are cut from the same plank with some variations: 45.6 $\pm$ 1.2mm, 10.1 $\pm$ 0.6 mm and 9.5 $\pm$ 0.7 mm (length, width, 
height, averaged over 25 samples). The compression, in the vertical direction and perpendicularly to the wood grains, is applied by a piston extension on 1cm$^2$ area of the sample top (Fig. 
\ref{fig:1}a, along the grain compressions were not tested). An Instron Electropuls E1000 testing machine is used to apply up to 1 kN at either constant strain- or stress-rate, of 0.10 
$\pm~0.01~\%$/s and 9 $\pm$ 1 kPa/s respectively, the variance in these (slow) rates being due to the different sample dimensions. To follow the microcrack dynamics, acoustic emission is 
detected with a piezoelectric transducer, attached on the sample upper side. The sampling frequency is 200 kHz. After thresholding, events are formed from data, when silent times shorter than 
35 $\mu\textup{s}$ are neglected, in standard practice. The event occurrence time t$_{i}$ is defined as the time of the event maximum amplitude and the energy $E_\text{i}$ is calculated as the 
square integral of the amplitude over the duration. Thus AE time series consist of pairs of inter-event waiting time and energy ($\tau_{i} = t_{i} - t_{i-1}$, $E_\text{i}$) and is analyzed with 
recent techniques. The number of events in a time interval divided by the interval length defines the event rate $r(t)$. Each sample produces typically a few thousand events: 4500 $\pm$ 2200 and 
5200 $\pm$ 3300 with constant strain- and stress-rates, respectively. The deformation field is captured by non-invasive digital image correlation (DIC) technique 
\cite{PapAE-DIC-11, DIC-2003, PaperDIC1-2010}. Pictures of one sample side are taken every second using a grey-scale camera (Dalsa Genie HM1024) and are then exploited in pairs to obtain the 
2-D local strain field \cite{PapAE-DIC-11, PaperDIC1-2010}.

\textit{Results -} Figures 1b-1d demonstrate that AE activity or avalanches are present in wood compression. This intermittency exists in all stages beyond the linear early response: consequent 
plastic deformation as indicated by a plateau in the stress-strain curve, and stiffening of wood, typical of cellular solids, are mixed with avalanche activity on all scales (Figure \ref{fig:1}b), 
to which question we return below for the consequences and origins of the varying event rate. For both loading protocols, no large events are visible initially, but the activity varies later, as 
expected from an avalanching system from small to very large events (Figure \ref{fig:1}c and 1d for interesting quantities: event energy, cumulative AE energy, event rate, and the cumulative number 
of events). Thus come three questions, two of them ubiquitous and one particular to wood: i) what are the statistical properties of the crackling noise? ii) what are the correlations in the AE 
signal like? iii) does the particular structure of wood become important or interact with the scale-free nature of acoustic emission or the deformation collective dynamics?

\begin{figure}
\includegraphics[width=\columnwidth]{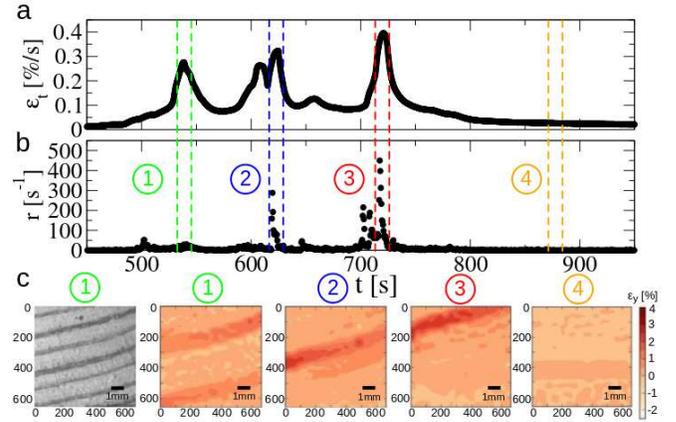}
\caption{\label{fig:2} a) Strain rate, calculated from sample height compression from testing machine and b) AE event rate, both for one constant stress-rate experiment; 4 particular instants 
numbered: 3 where high peaks in strain- and AE event-ratesare simultaneous, and one reference. c) Strain field from DIC for the 4 previously defined instants, with fixed color-scale for the vertical 
strain: the localization of deformation in softest layers occurs with high AE events- and strain-rates. The original image area is also shown (left).}
\end{figure}

\textit{Localization of deformation and AE - }
Wood mechanics is influenced by the multiscale structure, i.e the annual rings. High energy acoustic events are simultaneous with rapid reductions in sample height. Acoustic activity is also observed 
outside these periods of high strain rate, but the spikes in event rate correspond qualitatively well to those in strain rate. Figure \ref{fig:2} shows the mesoscopic strain rate $\epsilon_{t}$(t) 
(calculated from the decrease in time of the sample height, given by the displacement of the machine piston) and the event rate $r(t)$ to demonstrate their correlation with the localization of 
deformation from DIC \cite{Wood-2013}. The main observed phenomenon is the (partial) collapse of the annual rings \cite{Wood-2013}, leading to periods of high acoustic activity. Strain maps 
calculated by the DIC algorithm over particular 10 s time intervals (three over intervals of high acoustic activity and one reference) illustrate how the strain localizes at these events. The first 
shows that the strain is localized in two bands corresponding to annual rings. The same behavior appears in the next two strain maps, but is absent in the last. These structure-dependent correlations 
may be related to e.g. the clear precursor events of high energy seen right before the major collapses in silica-based porous samples \cite{SilicaCeramics-2014} and such rises in event rates are 
indicative of upcoming loss of structural stability \cite{Sethna-2001}.

\begin{figure}
\includegraphics[width=\columnwidth]{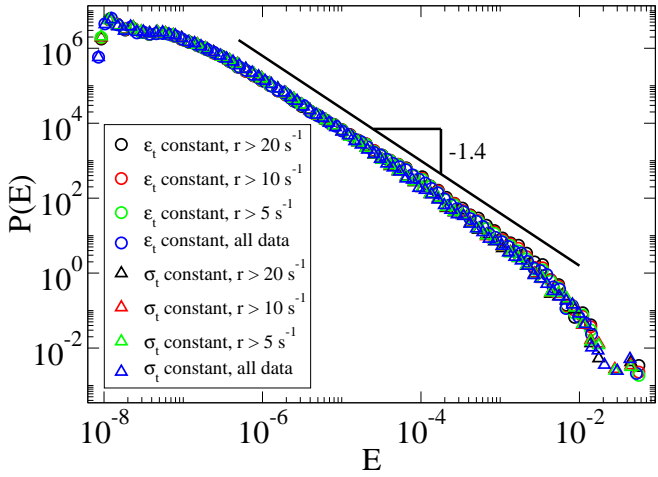}
\caption{\label{fig:3} Distribution function of energies of AE events, averaged over all the experiments realized at constant strain rate $\epsilon_{t}$ or stress rate $\sigma_{t}$ respectively, 
exhibit a power law $P(E) \sim E^{-\beta}$ with $\beta \approx 1.4$, the scaling region and exponent being almost identical in all experiments. Thresholding the data with the AE event rate $r$ 
has little effect on the distribution shape.}
\end{figure}

\textit{Statistical properties of acoustic emission - } 
AE data from wood compression turns out to follow several empirical deformation or seismological laws. The AE energies follows the Gutenberg-Richter (GR) law $P(E)\sim$E$^{-\beta}$ 
\cite{GR-1954, GR-1956} for both experiments with a constant strain- and stress-rate, with little difference: the $P(E)$ is well fitted, using maximum likelihood method, with a power law with 
exponent 1.37 $\pm$ 0.01 for $\epsilon_{t}$ constant and 1.43 $\pm$ 0.01 for $\sigma_{t}$ constant (Figure \ref{fig:3}). These  values are close to the ones found in brittle materials such as Vycor 
($\beta=$ 1.39 $\pm$ 0.02 \cite{Vycor-2011} or 1.40 $\pm$ 0.05 \cite{Vycor-2013} for tests at different stress rates), chipboard panels (1.51 $\pm$ 0.05 \cite{Precursors-1998}),  and for various 
silica-based porous materials ( $\beta$ between 1.35 and 1.55 \cite{SilicaCeramics-2014}). The exponent dependence on the fitting range is robust in a large region of around five decades (E from 
10$^{-7}$ to~10$^{-2}$), for both types of experiments, the only change being a very slight decrease in high energy event probability in constant strain rate experiments. In comparison to earthquake 
activity, the rate of acoustic events varies considerably over the experiment duration and therefore the process is non-stationary. This is taken into account by restricting the analysis of AE 
energy distributions to time intervals with event rate above some pre-determined threshold value, but these thresholds have almost no effect on the distributions (figure \ref{fig:3}). 

\begin{figure}
\includegraphics[width=\columnwidth]{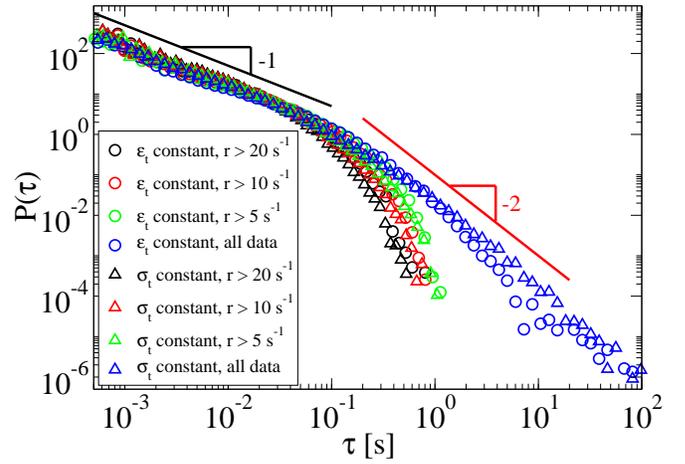}
\caption{\label{fig:4}  Waiting times of AE events are power law distributed for short waiting times ($\tau\approx$~10$^{-3}$-~10$^{-1}$s) with an exponent $\gamma\approx1$. Considering all data, 
a second scaling region for long waiting times is seen, with $\gamma\approx2$. Thresholding by the AE event rate $r$ introduces a cut-off to the distributions.}
\end{figure}
 
Here the waiting time distributions can be roughly divided into two parts depending on the loading protocol described by  two power laws with different exponents for long and short silent times 
(Figure \ref{fig:4}), arising from large variations in the event rate during an experiment (cf. the data thresholded with $r$ showing how a cut-off becomes visible as expected). At constant strain 
rate, the exponents are -0.97 $\pm$ 0.01 and -2.42 $\pm$ 0.03 for the two regimes divided with waiting times from 0.2 $\mu$s to 0.2 s and from 0.2 s to 2 s respectively. For constant stress rate, 
the first region is defined for waiting times from 0.01 s to 1 s with a power law exponent 1.03 $\pm$ 0.01 and the second exponent is 1.95 $\pm$ 0.02, for times between 1 s and 1000 s. The transition 
between regions is slightly smoother in the experiments with a constant stress rate. The difference in the tail exponents (about 2.4 vs. about 2) should be due to different event rate distributions 
in the two cases; in constant $\sigma_{t}$ case, the distribution also extends to larger waiting times. The fit with two power-laws with exponents around -1 and -2 for short and long waiting times 
respectively is analogous to other porous materials \cite{Vycor-2013, SilicaCeramics-2014}. The fat waiting time distributions imply temporal correlations in the deformation dynamics 
\cite{Alava-2006}, which we next study by using geophysics techniques and recent methods for laboratory experiments.

\begin{figure}
\includegraphics[width=\columnwidth]{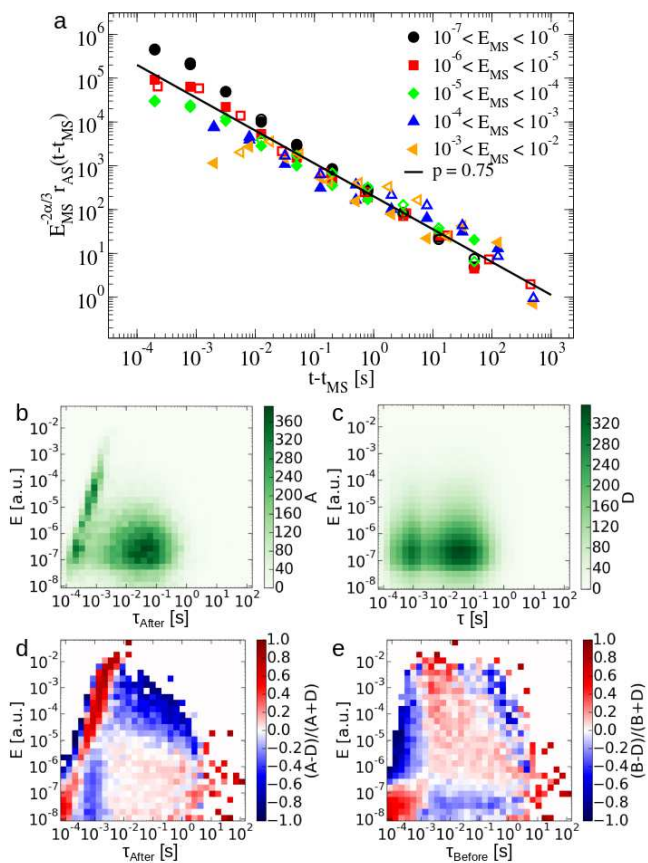}
\caption{\label{fig:5} a) Productivity law for different mainshock energy intervals. The scaled event rate decays with an exponent of 0.75. Empty symbols correspond to constant strain rate and 
filled symbols to constant stress rate.b) Number A of AE events for event energy against the waiting time after the considered event, in the constant strain rate case; the other one is quite 
similar. c) Number D of events for uncorrelated data, obtained by randomizing the ($\tau, E)$-pairs. d) The relative difference between the original A and randomized D data sets shows the presence 
of correlations. e) Relative difference when considering waiting times before an event, with B the number of these events.}
\end{figure}

\textit{Aftershocks and Omori's law - }
In the case of earthquakes, a mainshock is followed by smaller aftershocks, implying the famous seismology Omori's law \cite{Hirata-1987, Omori-1894, Utsu-1995, Helmstetter-03, PapAE-08}: the number 
of aftershocks per unit time interval $n(t)$ is $n(t)=K(t+c)^{-1}$ with K and c constants. The modified law relates the aftershock sequence to the time elapsed from the mainshock time \cite{PapAE-08}
$t_{MS}$ as a power law: $r_{AS}(t-t_{MS})~\propto~(t-\text{t}_{MS})^{-p}$ where $r(t)$ is the event rate of aftershocks and $p\sim1$. In fracture experiments, events of any energy could be 
considered as mainshocks to verify if such statistical laws hold \cite{Vycor-2013, Hirata-1987}.  Thus we vary the mainshock energy (threshold), so that all events between two subsequent mainshocks 
are aftershocks of the first one. The rates are then averaged over each mainshock energy interval. Finally these averaged rates are averaged over the different experiments. The event sequences 
exhibit an Omori's law which becomes less clear as the mainshock energy is increased. The Fig.~\ref{fig:5}a shows the productivity law \cite{Vycor-2013, Helmstetter-03}, which corresponds to the 
averaged Omori sequences where the aftershock event rate $r_{AS}$ is scaled by $E_{MS}^{-2 \alpha /3}$. The best data collapse is obtained for $\alpha \approx 0.4$. The plots for the constant stress 
rate experiments show a a more robust scaling (especially after $(t-t_{MS})$= 0.1 s). The event rate decays with an exponent around $p=0.75$, as in Vycor \cite{Vycor-2013}.

\textit{Event energy and waiting time correlations - }
As shown in Fig.~\ref{fig:5}b, the events are concentrated around certain values for interevent times and energies \cite{Vycor-2013}: here around ($\tau_{After},E$) = (0.04 s, $2\cdot10^{-7}$). 
In experiments at constant stress rate, this occurs at a slightly lower waiting time, but the features are otherwise similar. The waiting times after an event of a certain energy present a power 
law relation for aftershocks (high energies followed by short waiting times)  \cite{Helmstetter-03, PapAE-08}, which spans around five decades in energy with an exponent around 4. This relation is 
absent in the waiting time before an event. Some correlation between small energies and small waiting times is present instead. The comparison of the experiments at constant strain rate with an 
artificial uncorrelated dataset (Fig.~\ref{fig:5}c) (the ($\tau, E)$-pairs randomized, and repeated 100 times for better statistics) shows a gap with less events with energies in the most prominent 
range from 10$^{-7}$ to 10$^{-6}$ with waiting times from around 10$^{-3}$ to 10$^{-2}$ s. A similar but much smaller gap is found for constant stress rate. The relative differences between waiting 
times after an event and uncorrelated waiting times show that aftershock correlations exist (Fig.~\ref{fig:5}d), more clearly than in porous material like Vycor \cite{Vycor-2013}. The relative 
differences between the foreshock waiting times and uncorrelated waiting times show a correlation between small energies and small waiting times (Fig.~\ref{fig:5}e), and a slight correlation between 
high energy and high waiting time events across all energies and waiting times over 10$^{-3}$ s.

\textit{Conclusion - }
Wood compression shows clear evidence of scale-free avalanche activity, both in stress- or strain-rate controlled protocols. The statistics of the avalanches (distributions of AE events energies 
and waiting times) exhibit exponents reminiscent of brittle porous materials and geophysics. The dynamics of compression is correlated in further analogy with rocks and other such materials and 
indeed even earthquakes, as demonstrated by the modified Omori's law, the law of aftershock productivity and the universal scaling law for the waiting time distribution typically used in statistical 
seismology. Scale-free avalanches co-exist in wood in spite of the material structural features, clearly visible and relevant in compression curves or by microscopic analysis of the larger events 
(DIC). We conjecture that this fact is only manifested in terms of a varying event (or production) rate. Further studies would be needed to analyze whether softwood layer collapse is always 
correlated with an increased AE activity, and whether such large events are in fact scale-free or imply temporary changes in acoustic activity correlations. Further work is needed to advance models 
to reproduce these results, including also the wood cellular structure. Similarly, experiments are called for to explore the role of rheology (temperature \cite{Precursors-2002, Deschanel-2009, 
WoodAE-00} and humidity \cite{WoodAE-00, WoodMat-2011}) in wood and in other candidate materials of natural origin.


\vspace{0.5cm}
\textit{Acknowledgments} 
\\We acknowledge the financial support of the Academy of Finland through the COMP center of excellence (project no. 251748). 
\\

\end{document}